\documentclass[reprint, superscriptaddress, aps, prb, nofootinbib, balancelastpage]{revtex4-2}
\usepackage{graphicx} 
\usepackage{hyperref}
\usepackage{amsmath}
\usepackage{mathtools}
\usepackage{cancel}
\usepackage{placeins}
\usepackage{amssymb}
\usepackage{bm}
\usepackage{float}
\usepackage[T1]{fontenc}
\usepackage{mdframed}
\usepackage[normalem]{ulem}
\usepackage{acro}
\usepackage{chemformula}

\DeclareAcronym{DFT}{
short = DFT,
long = Density-Functional Theory
}

\DeclareAcronym{ACE}{
short = ACE,
long = Atomic Cluster Expansion
}

\DeclareAcronym{MACE}{
short = MACE,
long = Multi-ACE
}

\DeclareAcronym{MD}{
short = MD,
long = Molecular Dynamics
}

\DeclareAcronym{MLIP}{
short = MLIP,
long = machine-learned interatomic potential
}

\DeclareAcronym{PES}{
short = PES,
long = Potential Energy Surface
}

\hypersetup{
    colorlinks=true,
    linkcolor=blue,
    filecolor=magenta,      
    urlcolor=blue,
    citecolor=blue
    }

\def\R{\mathbb{R}}

\def\rr{{\bm r}}

\def\btheta{{\boldsymbol{\theta}}}
\def\blambda{{\mathbf{T}}}

\def\btheta{{\bm \theta}}

\newcommand{\FastDDD}{\texttt{FourierD3}}
\newcommand{\torchDDD}{\texttt{torch-dftd3}}

\begin{document}

\title{A fast summation method for the DFT-D3 dispersion correction}

\author{Victoria Valeeva}
\email{vicval@math.ubc.ca}
\author{Cheuk Hin Ho}
\affiliation{Department of Mathematics, University of British Columbia, 1984 Mathematics Road, Vancouver, BC, Canada V6T 1Z2}
\author{Mario Geiger}
\author{Franco Pellegrini}
\affiliation{NVIDIA, 2788 San Tomas Expressway Santa Clara, CA 95051, USA}
\author{G\'abor Cs\'anyi}
\affiliation{Max Planck Institute for Polymer Research, Ackermannweg 10, Mainz, 55128, Germany}
\affiliation{Engineering Laboratory, University of Cambridge, Trumpington Street, Cambridge, CB2 1PZ, UK}
\author{Emine Kucukbenli}
\affiliation{NVIDIA, 2788 San Tomas Expressway Santa Clara, CA 95051, USA}
\author{Christoph Ortner}
\affiliation{Department of Mathematics, University of British Columbia, 1984 Mathematics Road, Vancouver, BC, Canada V6T 1Z2}

\date{\today}

\begin{abstract}
The DFT-D3 dispersion correction is routinely added to machine learning force fields (MLFFs) trained on dispersion-deficient functionals such as PBE. Its environment-dependent pair coefficients, however, break the atom-centered separability that fast summation methods require, forcing practitioners either to truncate D3 or to accept a substantial slowdown. We introduce \FastDDD{}, a method that uses a functional low-rank decomposition to restore this separability and enable particle-mesh evaluation in $O(N\log N)$ time without a real-space cutoff on the dispersion sum.
\end{abstract}

\maketitle

\section{Introduction} 
Density functional theory (DFT) has long been the dominant computational framework for studying the electronic structure of molecules and materials, and a high-fidelity source of interatomic energies and forces. More recently, DFT has taken on a second role as the primary engine for generating training data for machine learning force fields (MLFFs)~\cite{drautz2019ace, batatia2022mace, lysogorskiy2026grace, batatia2025macemp, kovcs2025maceoff, musaelian2023allegro, batzner2022nequip, zubatyuk2019aimnet, anstine2025aimnet2}, where the ability to produce energy and force labels at scale is essential for building large and chemically diverse datasets. This role constrains the choice of DFT functional: dispersion-aware methods~\cite{tkatchenko2012mbd, becke2007exchangehole, tkatchenko2009ts, grimme2006semiempirical, grimme2010consistent, grimme2011effect, vdwdf, minnesota1, minnesota2, vv10} carry computational costs that are prohibitive at the millions-of-calculations scale typical of modern MLFF training sets. As a result, the workhorse functional for large-scale data generation has become PBE, a generalized gradient approximation (GGA) that is computationally efficient but lacks any description of long-range electron correlation. Systems in which these interactions are physically consequential --- e.g., biomolecular~\cite{piana2012bio1, norberg2000bio2, patra2003bio3, khabibrakhmanov2025bio4} and condensed-phase~\cite{linse1986liq1, sega2017liq2, gebbie2017liq3, schlaich2019liq4} systems --- are therefore systematically misdescribed at the level of the training data itself, a deficiency that propagates directly into any MLFF trained on it.

A parallel limitation is built into the models themselves: the most common MLFF architectures retain a local ansatz, decomposing the total energy into atomic contributions that depend only on features of an atom's local or semi-local environment. This locality is what allows MLFFs to scale linearly with system size and to evaluate forces in milliseconds per step, making molecular dynamics simulations accessible at length- and time-scales far beyond the reach of direct DFT.
A substantial body of recent work has been devoted to augmenting MLFFs with long-range electrostatic models~\cite{ko2021hdnnp, king2025les1, cheng2025les2, unke2021spookynet, huguenindumittan2023lode, batatia2026macepolar, khajehpasha2022cent, kocer2025behler}.

Long-range dispersion, in contrast, has received considerably less attention from the MLFF community~\cite{ying2023nep, tu2023neural, moerman2026mbdml}, despite being no less important physically: it is the dominant long-range attractive interaction in neutral systems and critically shapes the structure of liquids, molecular crystals, layered materials, and interfaces.

The canonical treatment of dispersion in MLFFs trained on PBE datasets is to add \textit{a posteriori} corrections originally designed for DFT energies and forces, as opposed to creating new datasets utilizing expensive non-local dispersion-inclusive electronic structure methods like vdW-DF \cite{vdwdf}.  Among these corrections, the DFT-D3 method~\cite{grimme2010consistent} has become a {\it de facto} standard. A distinguishing feature of D3 is its explicit dependence on the local atomic environment through coordination numbers, which allows the dispersion coefficients to respond to bonding and hybridization without any self-consistent calculation. In principle, D3 supplies precisely the non-local contribution that locally featurized MLFFs omit, without requiring additional training data.

In practice, coupling D3 to an MLFF comes with a high computational cost: because D3 is typically evaluated by direct real-space summation, obtaining its long-range contribution to within sub-meV accuracy requires constructing neighbour lists well beyond the native cutoff of the MLFF --- often 15~\AA\ for molecular liquids and in excess of 40~\AA\ for dense metallic systems, as we demonstrate in Sec.~\ref{sec:convergence}. The resulting neighbour list quickly dominates the cost of the simulation, nullifying the speed advantage that motivates the use of an MLFF in the first place. Current MLFF workflows therefore either truncate D3 at the MLFF cutoff \cite{ying2023nep}, absorbing an uncontrolled truncation error, or accept a substantial slowdown \cite{tu2023neural}; neither choice is satisfactory for production use.

For the analogous computational problem in electrostatics, fast summation techniques — Ewald summation and the particle-mesh methods that followed it (PME \cite{darden1993particle}, SPME \cite{essmann1995smooth}, P3M \cite{eastwood1984pmdpthe}) — have long provided an elegant solution, reducing the cost of a fully converged long-range interaction to $O(N \log N)$. The same methodology applies to any pairwise potential whose coefficients are separable across atoms, as is the case for classical point-charge electrostatics and for simple dispersion models such as DFT-D2 \cite{grimme2006semiempirical}, whose coefficients obey a geometric mixing rule. The environment dependence of D3, however, breaks this separability: the pair coefficient $C^{6}_{ij}(\theta_i, \theta_j)$ depends on the coordination numbers of both atoms in a non-multiplicative way, which precludes a direct application of particle-mesh methods.

In this work obstruction. We observe that, because the D3 pair coefficients depend on atomic environments through a smooth function of the coordination numbers, they admit an accurate low-rank tensor decomposition that can be computed once, offline, from the reference $C_6^{\text{ref}}$ tensor of Grimme et al.\ \cite{grimme2010consistent}. This decomposition separates the environment-dependent pair coefficient into a sum of products of atom-centered factors, each of which is then amenable to independent fast summation -- a similar matrix factorization approach in the case of static dispersion coefficients has been considered by Isele-Holder et al \cite{Isele-Holder2013-ou}. The resulting algorithm, which we call \FastDDD{}, evaluates the full D3 correction in $O(N \log N)$ time at a marginal cost that, for system sizes of practical interest, is dominated by the cost of the local neighbour list that the MLFF is already required to build. The dispersion sum itself is evaluated without any real-space cutoff; the only real-space cutoff that remains is the short coordination-number cutoff, which a classical or ML force field neighbour list already provides. Note that we only apply our approach to the pairwise dispersion terms, and, in particular, the optional ATM term of D3 is not considered in this work.

More generally, our approach introduces a mathematical framework --- a functional tensor decomposition of environment-dependent pair coefficients --- that is not specific to D3. The same method extends to more advanced dispersion models such as D4~\cite{caldeweyher2017extension, caldeweyher2019generally, caldeweyher2020extension}, XDM~\cite{becke2007exchangehole}, or Tkatchenko--Scheffler~\cite{tkatchenko2009ts}, although in those models the computational bottleneck instead lies in computing the atomic features --- electron densities or charges --- that enter the interaction model.

\section{Methods}

\subsection{Fast Summation}
Typically, the total energy of an atomistic system is first partitioned into short-range and long-range contributions,
\begin{equation}
    E_{\text{total}} = E_{\text{short}} + E_{\text{long}}.
\end{equation}
The short-range energy $E_{\text{short}}$ encompasses bonded interactions (e.g., covalent bonds, bond angles, and torsions) as well as steric repulsions. Long-range contributions $E_{\text{long}}$ include electrostatic (Coulomb) and van der Waals (dispersion) interactions.

The vast majority of long-range interaction models reduce to a pairwise sum of weighted contributions from a spherically symmetric potential $\varphi(r)$, as obtained from second-order perturbation theory~\cite{eisenschitz1930ber}. We consider a periodic system defined by a Bravais lattice $\Lambda$ with cell volume $\Omega$, containing $N$ atoms at positions $\rr_i \in \R^3$, elements $Z_i$, and weights $C_i(\btheta)$ depending on model-specific system parameters $\btheta$. The periodized energy is then given by 
\begin{equation}
    \label{eq:periodicE}
    \begin{aligned}
    E_\varphi = \sum_{\blambda\in \Lambda}\sum_{i,j=1}^NC_i(\btheta)C_j(\btheta)\varphi\left(|\rr_{ij}+\blambda|\right).
  \end{aligned}
\end{equation}
Assuming absolute integrability of $\varphi\left(|\rr|\right)$, we can apply Poisson's summation formula to rewrite
\begin{equation}
    \begin{aligned}
    E_\varphi = \frac{1}{\Omega}\sum_{\mathbf{k}\in \Lambda'}\sum_{i,j=1}^NC_i(\btheta)C_j(\btheta)\widehat{\varphi}\left(|\mathbf{k}|\right)e^{i\mathbf{k}\cdot\rr_{ij}}
  \end{aligned}
\end{equation}
where $\Lambda'$ is the dual of $\Lambda$. By defining the structure factor, we obtain the form of \eqref{eq:periodicE} common in the Ewald summation literature:
\begin{equation}
    \label{eq:energyEwald}
    \begin{aligned} 
        S(\mathbf{k}) 
        &= \sum_{i=1}^NC_i(\btheta)e^{i\mathbf{k}\cdot \rr_i}, 
        \\ 
        E_\varphi &= \frac{1}{\Omega}\sum_{\mathbf{k}\in\Lambda'}\widehat{\varphi}\left(|\mathbf{k}|)S(\mathbf{k}\right)S(\mathbf{-k}). 
    \end{aligned}
\end{equation}

In practice, $\varphi$ is typically very smooth, in which case reciprocal-space summation requires significantly smaller cutoffs than its real-space counterpart to achieve the same accuracy. Performance can be further improved by interpolating the complex exponentials $e^{i\mathbf{k}\cdot \rr}$ on an equispaced grid $\mathbb{M}$ and rewriting the structure factors as a Discrete Fourier Transform
\begin{equation}
    \begin{aligned}
        S(\mathbf{k}) \approx \widetilde{S}(\mathbf{k})&=\sum_{i=1}^NC_i(\btheta)\left[\sum_{\mathbf{x}\in\mathbb{M}}W_i(\mathbf{x})e^{i\mathbf{k}\cdot\mathbf{x}}\right]\\
        &=\sum_{x\in\mathbb{M}}\left[\sum_{i=1}^NC_i(\btheta)W_i(\mathbf{x})\right]e^{i\mathbf{k}\cdot\mathbf{x}}.
    \end{aligned}
\end{equation}
This is the approach taken by the family of particle-mesh methods: the Particle-Mesh-Ewald (PME) method~\cite{darden1993particle} uses Lagrange interpolation weights $W_i$. Both Smooth Particle-Mesh-Ewald (SPME)~\cite{essmann1995smooth} and Particle-Particle-Particle-Mesh (P3M)~\cite{eastwood1984pmdpthe} use Euler B-splines for interpolation; P3M additionally scales the potential $\varphi$ to correct for aliasing --- see~\cite{eastwood1984pmdpthe} for details. The decomposition developed below is agnostic to the choice of fast-summation backend: we use SPME in this work, but any of these particle-mesh schemes --- or the fast multipole~\cite{greengard1987fmm} and multilevel-summation~\cite{hardy2016multilevel} methods that also accommodate non-periodic and mixed boundary conditions --- may be substituted without modification.

\subsection{Tensor Decomposition}
\label{sec:tensordecomp}
Fast summation techniques rely on the assumption that the pairwise interaction $C_{ij}\varphi(r_{ij})$ can be decoupled into atom-centered contributions. In empirical force fields~\cite{walters2018evaluating} and simpler dispersion corrections such as DFT-D2~\cite{grimme2006semiempirical}, this separability is trivially achieved because the dispersion coefficients $C_{ij}$ are constant or satisfy simple mixing rules such as $C_{ij} = \sqrt{C_{ii}C_{jj}}$. Fast summation is then straightforward to apply, and is indeed already widely implemented in molecular simulation packages~\cite{abraham2015gromacs, LAMMPS, torchpme}.

In contrast, the more advanced DFT-D3 correction introduces a complex environment dependence: $C_{ij}$ depends on the coordination numbers of both atoms in a non-separable way; see Section~\ref{sec:D3} for details.

To apply fast summation methods to such a model, one must approximate the pairwise coefficients $C_{ij}$ by a separable low-rank sum
\begin{equation}
    C_{ij} \approx \sum_{\ell = 1}^{\ell_{\rm max}} \lambda_\ell C_{i,\ell} C_{j,\ell},
\end{equation}
where $\lambda_\ell$ is a scalar, resulting in an approximation of the energy: 
\begin{equation}
    \label{eq:lowrank_fastsum}
    \begin{aligned}
    E_\varphi \approx \sum_{\ell = 1}^{\ell_{\rm max}} \lambda_\ell \sum_{\blambda\in \Lambda}\sum_{i,j=1}^NC_{i, \ell}C_{j, \ell}\varphi\left(|\rr_{ij}+\blambda|\right).
  \end{aligned}
\end{equation}
With this form, the total energy can be evaluated by performing the reciprocal-space summations independently for each component indexed by $\ell$. A naive strategy to obtain these separable components would be to compute an eigenvalue decomposition (EVD) of the instantaneous $N \times N$ matrix of pairwise coefficients $C_{ij}$ at each MD timestep. This approach requires an expensive matrix factorization at every step, which would undermine the computational efficiency of the underlying fast summation.

To bypass the prohibitive cost of an on-the-fly EVD, we must exploit the mathematical structure of the dispersion model. The key observation is that, in all cases we are aware of, the pairwise coefficient is a continuous function of atomic coordination parameters, 
\begin{equation}
    C_{ij} = \mathcal{C}(\theta_{i},\theta_{j}).
\end{equation}
Rather than decomposing the instantaneous pair matrix at each step, we apply a functional tensor decomposition directly to $\mathcal{C}$, 
\begin{equation}
    \mathcal{C}(\theta,\theta')
    \approx \sum_{\ell} \lambda_\ell \mathcal{C}_\ell(\theta) \mathcal{C}_\ell(\theta'),
\end{equation}
which shifts the computational cost of the online EVD to a single offline tensor decomposition.

To illustrate the separability mechanism on a familiar closed form, consider the classical Eisenschitz--London model~\cite{eisenschitz1930ber}, where the pairwise coefficients take the form
\begin{equation}
    C_{ij} = \frac{3 \alpha_i^0 \alpha_j^0}{2} \frac{I_i I_j}{I_i + I_j},
\end{equation}
$\alpha_i^0$ is the static dipole polarizability and $I_i$ the atomic ionization potential. (This example exercises only the separability mechanism: the $I_i$ are constant throughout a simulation rather than environment-dependent. It is the non-separability of $I_i I_j / (I_i + I_j)$, not the environment dependence, that the decomposition resolves.) Since the factor $\frac{3}{2} \alpha_i^0 \alpha_j^0$ is already separable, we focus on the non-separable factor $I_i I_j / (I_i + I_j)$ and seek an approximate decomposition 
\[
    \frac{I_i I_j}{I_i + I_j}
    \approx I_{\rm max} \sum_{\ell = 1}^{\ell_{\rm max}} \lambda_\ell v_\ell(I_i/I_{\rm max}) v_\ell(I_j/I_{\rm max}).
\]
Taking $I_{\rm max} = 25\,{\rm eV}$ and $v_\ell$ to be polynomials on the interval $[0.125, 1]$ (the range of ionization potentials across the periodic table), one achieves relative errors of $2.7 \times 10^{-5}$ at rank $\ell_{\rm max} = 4$ and $5.6 \times 10^{-9}$ at rank $\ell_{\rm max} = 7$.

Crucially, this decomposition remains valid even if the $I_i$ are environment-dependent, as in the D3 (Section~\ref{sec:D3}), D4~\cite{caldeweyher2017extension, caldeweyher2019generally, caldeweyher2020extension}, XDM~\cite{becke2007exchangehole}, or Tkatchenko--Scheffler~\cite{tkatchenko2009ts} models.

\subsection{DFT-D3}
\label{sec:D3}
A key novelty of the DFT-D3 dispersion correction developed by Grimme et al.~\cite{grimme2010consistent} is the dependence of the pairwise dispersion coefficients on the local atomic environment, which is accomplished via atomic coordination numbers,
\begin{equation}
        \label{eq:D3:original_cn_fcn}
    \begin{aligned}
        R^{\text{cov}}_{ij}&=R_i^{\text{cov}}+R_j^\text{cov}\\
        \theta_{ij}(r) &= \begin{cases}
        \left[1 + \exp \left( -16 \left( \frac{4R_{ij}^{\text{cov}}}{3r} - 1 \right) \right) \right]^{-1} & r\leq R_{\text{cut}}\\
        0 &r > R_{\text{cut}}
        \end{cases}\\
        \theta_{i} &= \sum_{\blambda\in\Lambda}{\sum_{j}}^{\prime} \theta_{ij}(|\rr_{ij}+\blambda|)
    \end{aligned}
\end{equation}
where $R_i^{\text{cov}}$ are covalent radii from Pyykk\"{o} et al.\ \cite{pyykk2008molecular}, $R_{\text{cut}}$ is the neighbour list cutoff radius, and $\sum'$ indicates omission of the self-interaction $j=i$. 

A reference dataset of $C^6$ dispersion coefficients is then constructed through a modified Casimir--Polder expression
\begin{equation}
    \begin{aligned}
        & C^{6, \text{ref}}_{Z_iZ_j}\left(\theta^{\text{ref}, p}_{Z_i}, \theta^{\text{ref}, q}_{Z_j}\right)  
        \\ 
        &=
        \frac{3}{\pi}\int_0^\infty\frac{1}{m}\left[\alpha_{(Z_i)_mH_n}(iw)-\frac{n}{2}\alpha_{H_2}(iw)\right] 
        \\ 
        &\qquad \times \frac{1}{k}\left[\alpha_{(Z_j)_kH_l}(iw)-\frac{l}{2}\alpha_{H_2}(iw) \right]\, dw, 
    \end{aligned}
\end{equation}
where $\alpha$ are dynamic polarizabilities of model hydrides with different coordination environments (indexed by $p$ and $q$ above), obtained via TD-DFT; see~\cite{grimme2010consistent} for details. The pairwise dispersion coefficients for a given system are then constructed by interpolating over the reference coordination numbers $\theta_{Z}^{\text{ref}, p}$,
\begin{equation}
\label{eq:coeffs}
    \begin{aligned}
        L_{i}^{p} &= \exp\left(-4 \big|\theta_i - \theta_{Z_i}^{\text{ref}, p}\big| ^2\right), \\ 
        C^6_{ij}(\theta_i, \theta_j)&=\frac{\sum_{p, q}C^{6, \text{ref}}_{Z_iZ_j}(\theta^{\text{ref}, p}_{Z_i}, \theta^{\text{ref}, q}_{Z_j})L_{i}^p L_{j}^{q}}{\sum_{p, q} L_{i}^p L_j^q}.
    \end{aligned}
\end{equation}
The dipole-quadrupole coefficients $C^8$ are estimated via a recursive relation:
\begin{equation}
    \begin{aligned}
    Q_{Z} &= \frac{1}{2}\sqrt{Z}\frac{\langle r^4 \rangle_Z}{\langle r^2\rangle_Z},\\
    C^8_{ij}(\theta_i, \theta_j) &= 3C^6_{ij}(\theta_i, \theta_j)\sqrt{Q_{Z_i}Q_{Z_j}},
    \end{aligned}
\end{equation}
where $\langle r^4 \rangle_Z$ and $\langle r^2 \rangle_Z$ are quadrupole- and dipole-moment expectation values derived from atomic densities~\cite{grimme2010consistent}.
The D3 energy correction is then computed via
\begin{align}
    &E_{\rm D3} = \\ 
    \notag 
    &-\frac{1}{2}\sum_{n=6,8}\sum_{\blambda\in\Lambda}\sum_{i,j}^{'}\frac{C^n_{ij}(\theta_i, \theta_j)}{|\rr_{ij}+\blambda|^n}f_{n}^{\text{damp}}\big(|\rr_{ij}+\blambda|, Z_i, Z_j\big),
\end{align}
where $f_n^{\text{damp}}$ is a damping function that prevents double-counting of short-range dispersion with the exchange-correlation functional. The most popular choices are the Becke--Johnson damping
\begin{equation}
    \begin{aligned}
        R_0(Z_i, Z_j) &= \sqrt{3\sqrt{Q_{Z_i}Q_{Z_j}}}, \\
        f_n^{\text{damp}}\left(r, Z_i, Z_j\right)&=\frac{s_nr^n}{r^n+\left(a_1R_0(Z_i, Z_j)+a_2\right)^n}
    \end{aligned}
\end{equation}
and the zero-damping
\begin{equation}
    \begin{aligned}
        f_n^{\text{damp}}\left(r, Z_i, Z_j\right) = \frac{s_n}{1+6\left(\frac{r}{s_{R, n}R_0(Z_i, Z_j)}\right)^{-\alpha_n}}, 
    \end{aligned}
\end{equation}
where $s_6 = 1, \alpha_6 = 14, \alpha_8 = 16, s_{R, 8} = 1$ are fixed, while $a_1, a_2, s_8, s_{R, 6}$ are adjustable empirical parameters fitted to a given exchange-correlation functional.

Our implementation of the D3 dispersion correction introduces one important modification. The D3 coordination number function~\eqref{eq:D3:original_cn_fcn}
has a non-zero limit as $r_{ij} \to \infty$, which makes the choice of a coordination number cutoff arbitrary.
Increasing the coordination number cutoff therefore results in a spurious energy limit.
We defer a detailed discussion and our proposed modification to Appendix~\ref{sec:app_divergence_cn}.
All results below employ a modified coordination number function that reproduces the standard D3 implementation to high accuracy, but converges smoothly to zero at $R_{\rm cut}$: 
\begin{align}
    \notag 
    R^{\text{mid}}_{ij}&=\frac{1}{2}\left(R^{\text{cov}}_{ij}+R_{\text{cut}}\right), 
    \\
    \label{eq:D3:smooth_cn_function}
    t_{ij}(r) &= \begin{cases}
        16+\left( \frac{r - R^{\text{mid}}_{ij}}{R_{\text{cut}} - r} \right)^2 \hspace{-0.5em}, & r > R_{ij}^{\text{mid}} \\ 
        16, &r \leq R_{ij}^{\text{mid}}
    \end{cases}
    \\
    \notag 
    \theta_{ij}(r) &= \begin{cases}
        \left[1 + \exp \left( -t_{ij}(r) \left( \frac{4R_{ij}^{\text{cov}}}{3r} - 1 \right) \right) \right]^{-1} \hspace{-1em},  & r\leq R_{\text{cut}}\\
        0, &r > R_{\text{cut}}
    \end{cases}
    \\
    \notag 
    \theta_{i} &= \sum_{\blambda\in\Lambda}\sum_{j}^{'} \theta_{ij}(|\rr_{ij}+\blambda|). 
\end{align}

Throughout, we use $R_{\text{cut}} = 6$~\AA, since it is a common cutoff for foundation MLFFs such as MACE-MP~\cite{batatia2025macemp}. For a median covalent radius of $1.31$~\AA, this results in errors of $\|\theta_{ij}^{\text{\FastDDD{}}}-\theta_{ij}^{\text{DFT-D3}}\|_{\infty}=0.0025$ and $\| \frac{d}{dr}\theta_{ij}^{\text{\FastDDD{}}}-\frac{d}{dr}\theta_{ij}^{\text{DFT-D3}}\|_{\infty}=0.0020$ compared to the standard CN function with the default CN cutoff of $20$~\AA.

\subsection{The \FastDDD{} method}
To apply the fast summation technique to DFT-D3, we approximate the $C^6_{ij}$ coefficients via a low-rank decomposition as described in Section~\ref{sec:tensordecomp}. We note that $C^{6, \text{ref}}$ is a rank-$4$ tensor with the symmetry
\begin{equation}
    C^{6, \text{ref}}_{Z_i, Z_j} = \left(C^{6, \text{ref}}_{Z_j, Z_i}\right)^T
\end{equation}

Restricting $Z$ to a set $\mathcal{Z}$ of elements present in the system under consideration, we define a symmetric block matrix
\begin{equation}
    M =\begin{bmatrix}
        C^{6,\text{ref}}_{11} & C^{6, \text{ref}}_{12} & \dots & C^{6,\text{ref}}_{1|\mathcal{Z}|} \\
        (C_{12}^{6,\text{ref}})^T& \ddots & & \vdots \\
        \vdots & & \ddots & \vdots \\
        (C_{1|\mathcal{Z}|}^{6,\text{ref}})^T & \dots & & C^{6,\text{ref}}_{|\mathcal{Z}||\mathcal{Z}|}
    \end{bmatrix}
\end{equation}
and compute the eigendecomposition with the Lanczos algorithm~\cite{lanczos1950iteration, 2020SciPy-NMeth} to the desired accuracy:
\begin{equation}
\label{eq:decomp}
    C^{6, \text{ref}}_{Z_iZ_j}(\theta^{\text{ref}, p}_{Z_i}, \theta^{\text{ref}, q}_{Z_j})=\sum_{\ell}\lambda_\ell v_{\ell, Z_i}\left(\theta^{\text{ref}, p}_{Z_i}\right)v_{\ell, Z_j}\left(\theta^{\text{ref}, q}_{Z_j}\right).
\end{equation}
The pairwise coefficients $C^6_{ij}(\theta_i, \theta_j)$ are then decomposed as
\begin{equation}
    \begin{aligned}
        C^6_{\ell, i}(\theta_i)&=\frac{\sum_p v_{\ell,Z_i}(\theta_{Z_i}^{\text{ref}, p})L_i^p}{\sum_p L_i^p}\\
        C^6_{ij}(\theta_i, \theta_j)&=\sum_{\ell}\lambda_\ell C^6_{\ell, i}(\theta_i)C^6_{\ell, j}(\theta_j),
    \end{aligned}
\end{equation}
giving the energy expression
\begin{equation}
\begin{aligned}
    E_{\rm D3}=&-\frac{1}{2}\sum_{\ell}\sum_{X, Y \in \mathcal{Z}}\sum_{\blambda\in\Lambda}\sum_{Z_i = X, Z_j = Y}\lambda_\ell \\
        & \qquad  \times C^6_{\ell, i}(\theta_i)C^6_{\ell, j}(\theta_j)\varphi_{X, Y}(|\rr_{ij} + \blambda|)\\
    &+V_{\text{self}}, 
\end{aligned}
\end{equation}
where the potential is
\begin{equation}
    \varphi_{X, Y}(r)=\frac{f_6^{\text{damp}}(r, X, Y)}{r^6}+\frac{3\sqrt{Q_XQ_Y}f_8^{\text{damp}}(r, X, Y)}{r^8}
\end{equation}
and the self-interaction term evaluates to
\begin{equation}
    V_{\text{self}}=\lim_{r\to 0^{+}}\frac{1}{2}\sum_{\ell}\sum_{X\in \mathcal{Z}}\sum_{Z_i = X}\lambda_\ell\left(C^6_{\ell,i}(\theta_i)\right)^2\varphi_{X, X}(r).
\end{equation}
Applying the theory of fast summation, we finally obtain
\begin{align}
    \label{eq:final_energy}
    \notag 
    S_{X}^\ell(\mathbf{k})&=\sum_{Z_i = X}C_{\ell,i}^6(\theta_i)e^{i\mathbf{k}\cdot
        \rr_i}\\
    E_{\rm D3}&=V_{\text{self}} 
    \\ 
    \notag 
            & -\frac{1}{2\Omega}\sum_{\ell}\sum_{\substack{X\in \mathcal{Z} \\ Y\in \mathcal{Z}}}\sum_{\mathbf{k}\in\Lambda'}\lambda_\ell\widehat{\varphi}_{X, Y}(|\mathbf{k}|)S_{X}^\ell(\mathbf{k})S_{Y}^\ell(-\mathbf{k}),
\end{align} 
where the Fourier transform of this potential can be derived analytically via contour integration.

For systems with a very large number of different chemical elements $\mathcal{Z}$, the double-sum over element types may become prohibitive. In this case one can apply a further functional tensor decomposition to $f_n^{\rm damp}$ and absorb $\sum_{X, Y}$ into $\sum_r$. This increases the rank of the decomposition but avoids the quadratic scaling in chemical complexity. An outline of the procedure is provided in Appendix~\ref{sec:decomp_damp}, which shows that the rank depends on the choice of damping-function parameters and the desired tolerance, but is independent of the number of chemical elements. This decomposition, however, introduces a non-trivial Fourier transform of the potential, which precludes a fast practical implementation of the method.

\section{Results}
We report the results of two independent implementations: \FastDDD{}{\tt -torch} implemented in PyTorch \cite{pytorch}, and \FastDDD{}{\tt -acc}, with accelerated CUDA kernels. In the main text, all timing results except those for the \ch{SiO_2} polymorphs (Sec. \ref{sec:sio2_ranking}) are measured using \FastDDD{}{\tt -acc}. Accuracy results are nearly indistinguishable between the two implementations and are therefore reported under the label "\FastDDD{}". A direct comparison of performance and accuracy between the two implementations is provided in Figure \ref{fig:convergence_fp32}.

\subsection{Efficiency of Tensor Decomposition}
\label{sec:results:decomp}
The computational efficiency of the \FastDDD{} framework is determined by the rank of the low-rank approximation applied to the $C^{6,{\rm ref}}$ reference tensor, since this rank dictates the number of reciprocal-space evaluations required (see Eq.~\eqref{eq:final_energy}). Truncating the sum in Eq.~\ref{eq:decomp} at rank $R$, we measure the accuracy of the decomposition by the maximum relative error
\begin{equation}
\begin{aligned}
    &\varepsilon_{\max}(R) = \\ 
    &\max_{Z_i, Z_j, p, q}
    \left| 1 - \frac{\displaystyle\sum_{\ell=1}^{R} \lambda_\ell\,
        v_{\ell,Z_i}\!\left(\theta^{\mathrm{ref},p}_{Z_i}\right)
        v_{\ell,Z_j}\!\left(\theta^{\mathrm{ref},q}_{Z_j}\right)}
        {C^{6,\mathrm{ref}}_{Z_i Z_j}\!\left(\theta^{\mathrm{ref},p}_{Z_i},
        \theta^{\mathrm{ref},q}_{Z_j}\right)} \right| ,
\end{aligned}
\end{equation}
the maximum running over all element pairs $Z_i, Z_j \in \mathcal{Z}$ and reference environments $p, q$ for which the reference coefficient is non-zero. To assess how this rank scales with chemical complexity, we computed the minimum rank needed to achieve a maximum relative error below $0.01\%$, as a function of the number of unique elements in the system. We adopt this threshold as the default tolerance in our implementation, since it balances accuracy against decomposition rank for conventional systems with fewer than 10 distinct elements.

\begin{figure}[htbp!]
    \centering
    \includegraphics[width=0.4\textwidth]{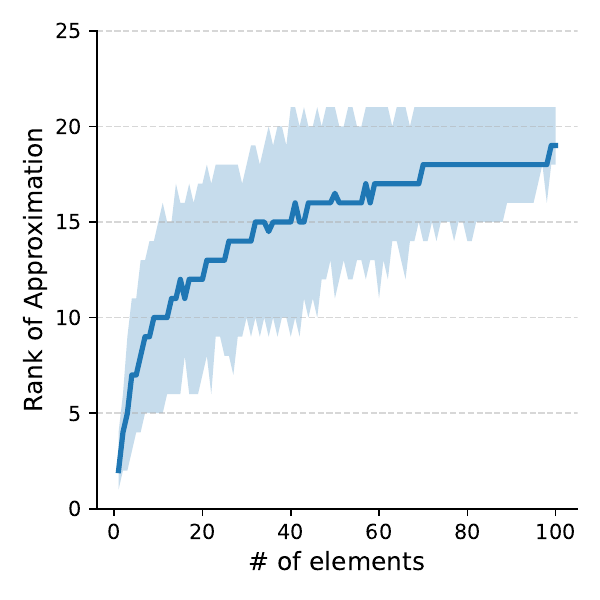}
    \caption{Median rank required to approximate the DFT-D3 $C^{6,{\rm ref}}$ reference tensor with maximum relative error below $0.01\%$, as a function of the number of unique atomic elements. The shaded region shows the min--max range across 100 randomly sampled element combinations for each element count.}
    \label{fig:decomp_rank}
\end{figure}

Figure~\ref{fig:decomp_rank} shows the median rank, together with the min--max range, across 100 random element combinations for subset sizes up to 100 elements. The DFT-D3 reference coefficients are strongly linearly dependent, yielding a highly compressible tensor. We observe that the required rank scales sublinearly with chemical complexity: for conventional systems with fewer than 10 distinct elements, the decomposition converges at a rank below 10.

Crucially, the rank saturates at 21 even for systems with up to 100 distinct elements. Since the rank remains small and bounded, the tensor decomposition keeps long-range dispersion evaluation at a small fraction of the cost of direct summation.

\subsection{Convergence and Consistency}
\label{sec:convergence}
We validate the consistency of the \FastDDD{} framework against the widely used \torchDDD{} library~\cite{takamoto2021torchd3}, which relies on direct real-space summation. Both methods use the same modified coordination-number function (Section~\ref{sec:D3}; Appendix~\ref{sec:app_divergence_cn}), so the comparison isolates the summation error.
Numerical experiments were performed on three chemically distinct periodic systems, each containing approximately 1000 atoms: liquid water, liquid benzene, and a 7-element high-entropy alloy (HEA: Al-Cu-Ag-Au-Ni-Pd-Pt). These systems were chosen to span a range of densities and chemical compositions representative of common simulations. For water and benzene, standard densities of $1.0$ and $0.88$~g/cm$^3$ were used. The HEA structures were generated with an fcc lattice constant taken as the average of the representative constants of the seven elements.
To establish a ground truth, reference energies and forces were computed using a \FastDDD{} evaluation, converged to within a stringent tolerance: a large reciprocal-space cutoff $k_{\text{cut}} = 9.0$~\AA$^{-1}$ and tensor-decomposition ranks chosen to bring the $C_6$ approximation error to machine precision.

\begin{figure*}[tp!]
    \centering
    \includegraphics[width=0.8\textwidth]{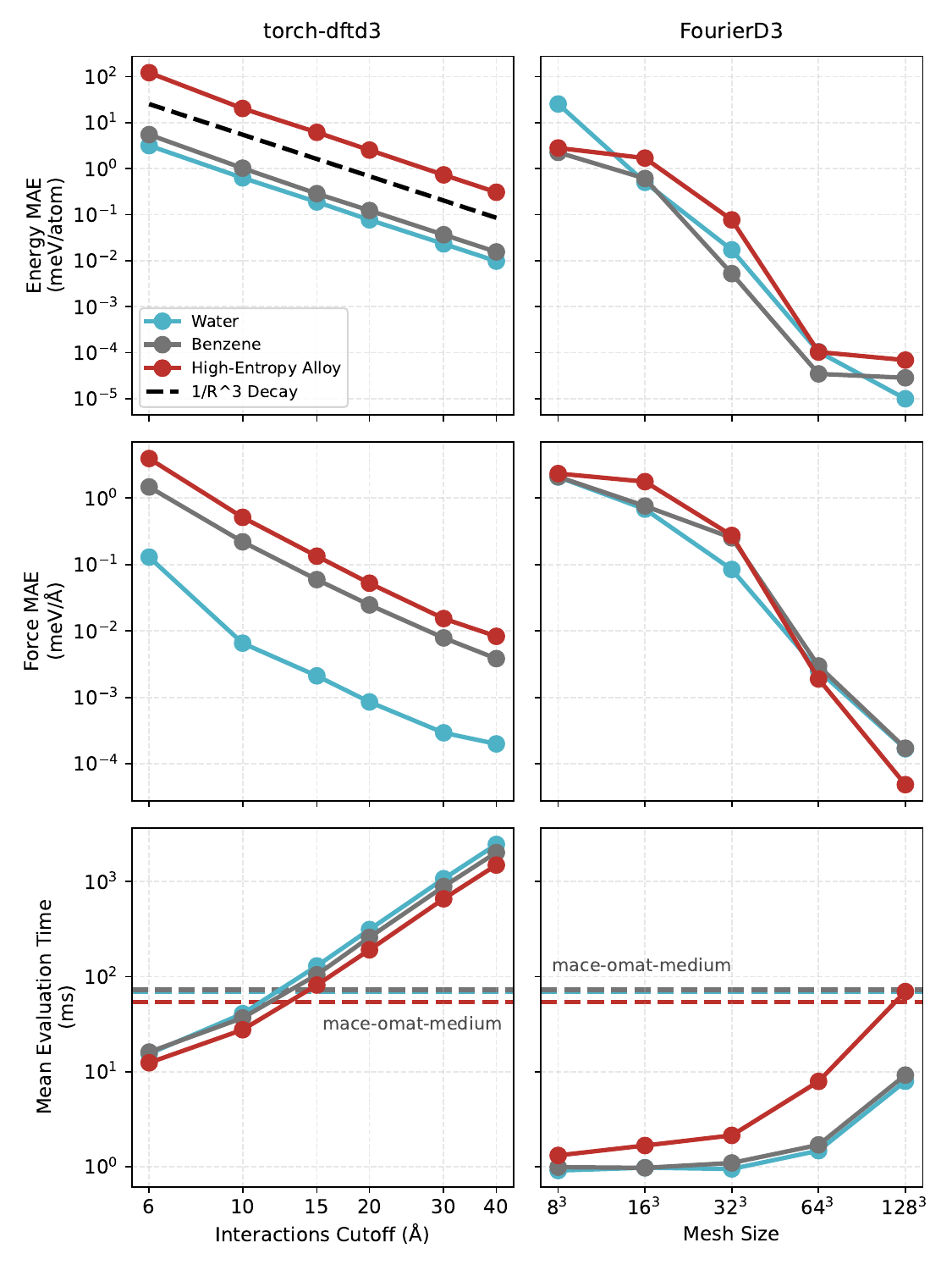}
    \caption{
    \textbf{Convergence and computational cost of DFT-D3 evaluation methods.}
    Mean Absolute Error (MAE) of predicted energies (top row) and atomic forces (middle row), evaluated against a converged \FastDDD{} reference ($k_{\rm cut} = 9.0$~\AA$^{-1}$), and mean evaluation time (bottom row). Performance is compared across three periodic systems of approximately 1000 atoms: liquid water (blue), liquid benzene (grey), and a 7-element HEA (red). Dashed horizontal lines in the bottom panels indicate the mean evaluation time of the \texttt{mace-omat-medium} architecture (executed with \texttt{cuEquivariance} in \texttt{Double} precision) for the corresponding system.
    \textbf{Left column:} Real-space summation (\torchDDD{}) as a function of the interaction cutoff radius. The dashed black line in the top-left panel illustrates the theoretical $\mathcal{O}(1/r^3)$ decay of the energy truncation error for a $1/r^6$ potential. Evaluation time scales cubically ($\mathcal{O}(r^3)$) with the interaction cutoff.
    \textbf{Right column:} SPME-based \FastDDD{} implementation as a function of the reciprocal mesh size, with a fixed 6.0~\AA\ coordination number cutoff. \FastDDD{} converges algebraically to the converged \FastDDD-Ewald reference in \texttt{Double} precision. The plotted evaluation times strictly exclude the 6.0~\AA\ neighbour list construction, representing the marginal cost of integration with an MLFF that natively computes this list.
    }
    \label{fig:convergence}
\end{figure*}

The left column of Fig.~\ref{fig:convergence} shows the convergence and computational cost of direct summation (\torchDDD{}) as a function of the real-space interaction cutoff. Because the DFT-D3 pairwise potential decays as $1/r^6$, integrating this interaction over a 3D volume yields a truncation error decaying as $1/r^3$. Consequently, achieving accuracy of sub-meV per atom requires extending the neighbour list cutoff to at least 15~\AA\ for water and benzene, and beyond 40~\AA\ for the denser HEA. As the summation volume grows, the evaluation time scales cubically ($\mathcal{O}(r^3)$), quickly becoming the bottleneck of the simulation. At an interaction cutoff of 40~\AA, a single evaluation step reaches approximately $10^3$~ms, negating the speed advantage of modern MLFFs.

By mapping the long-range interactions to a reciprocal-space grid, \FastDDD{} (right column of Fig.~\ref{fig:convergence}) avoids the high cost of direct summation. For standard meshes ($8^3$ to $32^3$), the \FastDDD{} evaluation time remains flat at under 10~ms. In the reported implementation, the charge-spreading operation is dominated by the number of particles rather than the mesh size. Overhead begins to grow at a mesh size of $64^3$, where the cost of the 3D Fast Fourier Transform (FFT) becomes dominant.

In many scenarios a neighbour list is already available. The \FastDDD{} timings we report therefore exclude the cost of constructing the 6.0~\AA\ neighbour list. The evaluation times for the \texttt{mace-omat-medium} model \cite{mace_foundations_2024, maceomat0} are shown for comparison. Because contemporary MLFFs natively compute local neighbour lists (typically spanning 5--6~\AA) to encode local atomic environments, the neighbour list required by \FastDDD{} comes effectively for free during a standard molecular dynamics step, since the same list can be reused.

\begin{figure*}[htbp]
    \centering
    \includegraphics[width=0.95\textwidth]{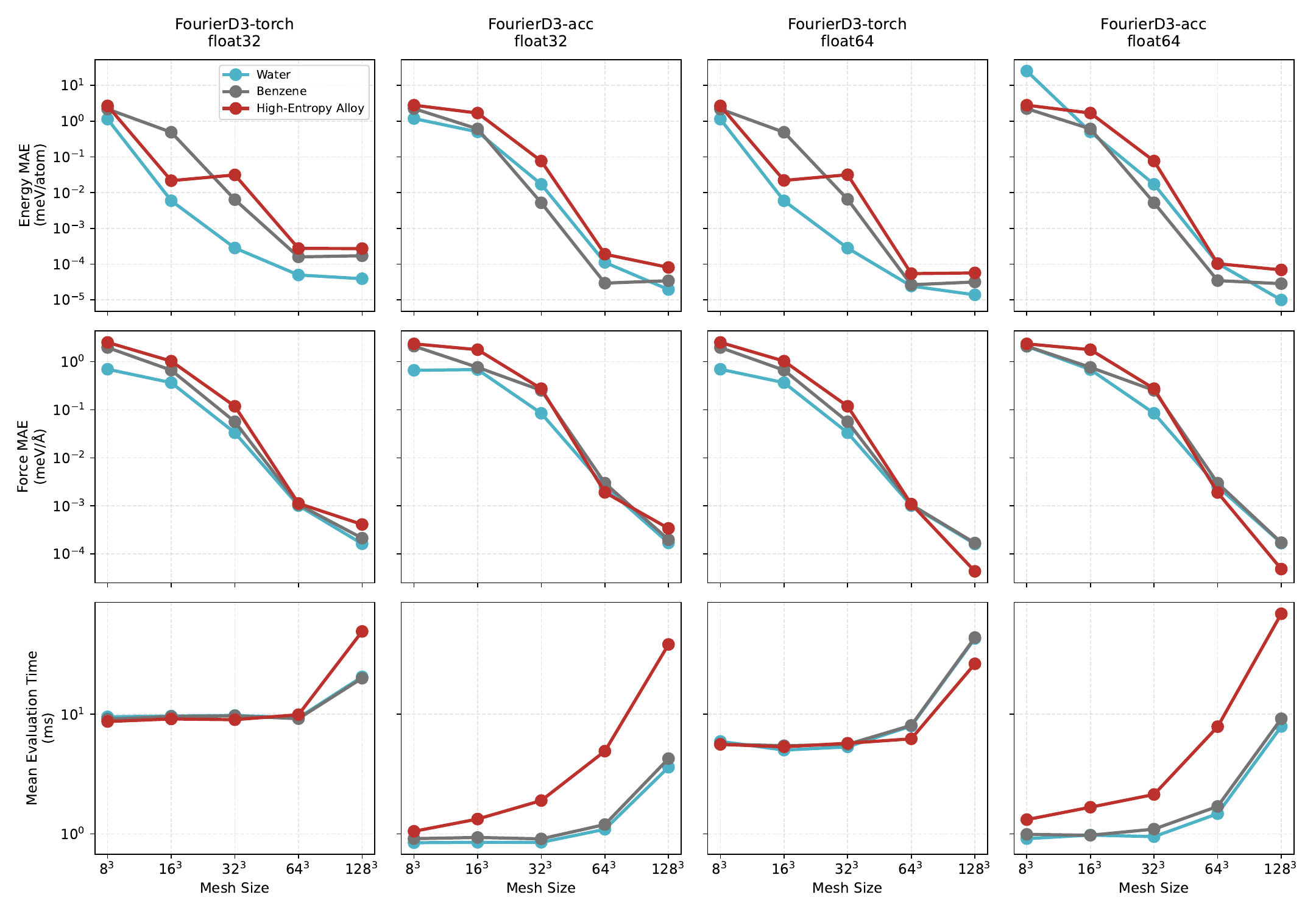}
    \caption{
    \textbf{\FastDDD{}{\tt -torch} and \FastDDD{}{\tt -acc} implementations in \texttt{Single} and \texttt{Double} precision.}
    Mean Absolute Error (MAE) of predicted energies (top row) and atomic forces (middle row), evaluated against a tightly converged Ewald reference ($k_{\rm cut} = 9.0$~\AA$^{-1}$), and mean evaluation time (bottom row). Performance is compared across three periodic systems of approximately 1000 atoms: liquid water (blue), liquid benzene (grey), and a 7-element HEA (red).
    }
    \label{fig:convergence_fp32}
\end{figure*}

Running \FastDDD{} in \texttt{Single} precision marginally changes the timings and accuracy - see Fig.~\ref{fig:convergence_fp32} for details.

\subsection{Scaling Performance}

We evaluate the scaling of \FastDDD{} as a function of system size. Benchmarks are performed on liquid water, liquid benzene, and the 7-element HEA, with periodic systems of 100 to 25,000 atoms (see Sec.~\ref{sec:convergence} for details). To maintain consistent accuracy as the system grows, the mesh is enlarged accordingly, as summarized in Table~\ref{tab:mesh_size}.

\begin{table}[htpb]
    \centering
    \caption{Reciprocal-space mesh size as a function of system size $N$.}
    \label{tab:mesh_size}
    \begin{tabular}{cc}
        \hline
        \textbf{System Size ($N$)} & \textbf{Mesh Size} \\ 
        \hline
        $N \le 200$ & $16^3$ \\
        $200 < N \le 2000$ & $32^3$ \\
        $2000 < N \le 20{,}000$ & $64^3$ \\
        $N > 20{,}000$ & $128^3$ \\ 
        \hline
    \end{tabular}
\end{table}

\begin{figure}[htbp!]
    \centering
    \includegraphics[width=\linewidth]{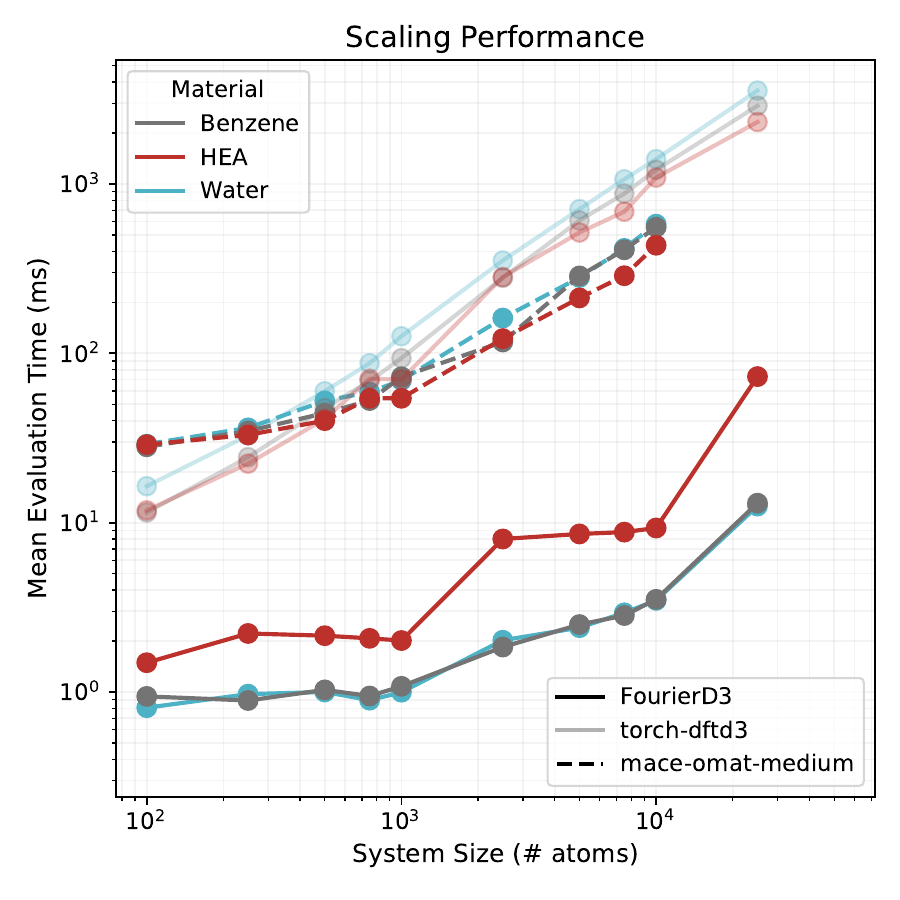}
    \caption{Evaluation time of \FastDDD{} as a function of system size $N$ across liquid water (blue), liquid benzene (grey), and HEA (red). Dark solid lines denote \FastDDD{}, faded solid lines denote \texttt{torch-dftd3} (6.0~\AA\ CN cutoff, 15.0~\AA\ interaction cutoff), and dashed lines denote \texttt{mace-omat-medium}. \FastDDD{} timings exclude the cost of neighbour list construction. The mesh size is scaled with $N$ to maintain numerical accuracy (cf.\ Table~\ref{tab:mesh_size}).}
    \label{fig:scaling}
\end{figure}

Mean evaluation times are shown in Fig.~\ref{fig:scaling}. To put the overhead in context, we compare \FastDDD{} timings against those of \texttt{mace-omat-medium} and \texttt{torch-dftd3} (using a 6.0~\AA\ coordination number cutoff and a 15.0~\AA\ interaction cutoff). Excluding the neighbour list construction --- the true marginal cost when \FastDDD{} is integrated with an MLFF that natively computes local neighbour lists --- the \FastDDD{} evaluation time remains flat at 1--2~ms for systems up to 10,000 atoms. This plateau corresponds to meshes up to $64^3$, for which the reciprocal-space summations are highly efficient on the GPU. Crucially, \FastDDD{} remains substantially faster than the MACE baseline and \torchDDD{} across all tested sizes, demonstrating that it can be coupled with modern MLFFs at negligible overhead.

\subsection{\ch{SiO2} Polymorph Ranking}
\label{sec:sio2_ranking}

Next, we illustrate the importance of avoiding a real-space interaction cutoff in three application scenarios. 
In our first example, we examine the relative stability ranking of silicon dioxide (\ch{SiO2}) polymorphs, demonstrating how strongly phase ordering depends on long-range dispersion. \ch{SiO2} has a rich phase landscape, making it an ideal test case for the influence of dispersion beyond the short- and medium-range regime. Our goal here is not to benchmark the accuracy of PBE or D3, but to highlight the sensitivity of polymorphic systems to truncation of long-range dispersion.

We retrieved all experimentally observed \ch{SiO2} structures with available PBE energy data from the Materials Project~\cite{mp1, mp2} (full list of MP-IDs in Appendix~\ref{sec:sio2_data}). Since an experimental stability ranking is not available, to establish the notion of ground truth, we augmented PBE energies with a tightly converged \FastDDD{} dispersion correction ($k_{\rm cut} = 30.0$~\AA$^{-1}$). Accuracy is quantified by counting pairwise ranking errors --- pairs of polymorphs whose predicted order disagrees with the reference.

\begin{figure}[htbp!]
    \centering
    \includegraphics[width=\linewidth]{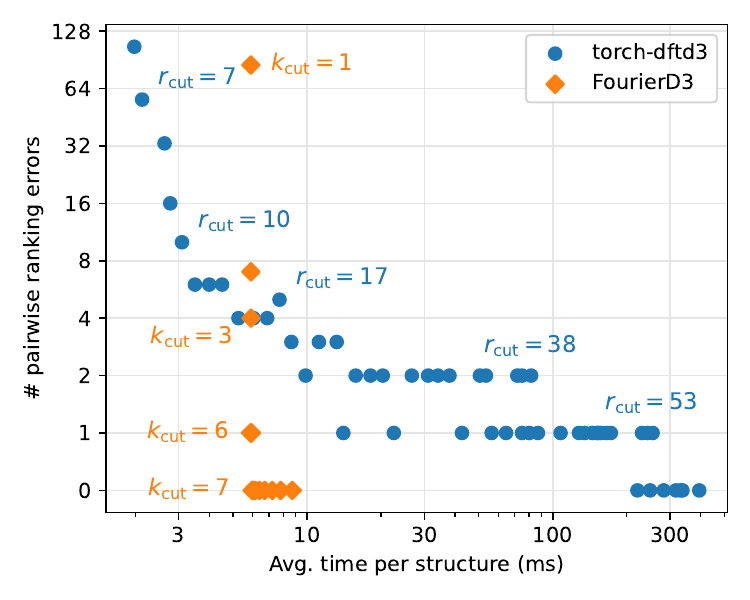}
    \caption{Pairwise ranking errors in the predicted stability ranking of experimentally observed \ch{SiO2} polymorphs, as a function of dispersion cutoff. Base PBE energies are retrieved from the Materials Project~\cite{mp1, mp2}. The reference ranking augments these with a tightly converged \FastDDD{} correction ($k_{\text{cut}}=30.0$~\AA$^{-1}$). The left panel shows the convergence of direct real-space summation (\torchDDD{}), which requires a 54~\AA\ interaction cutoff to eliminate ranking errors. The right panel shows the rapid convergence of \FastDDD{}, with zero ranking errors at a reciprocal cutoff of $7$~\AA$^{-1}$. Note that the real-space cutoff $r_{\rm cut}$ (left panel) is given in \AA, while the reciprocal cutoff $k_{\rm cut}$ (right panel) is given in \AA$^{-1}$.}
    \label{fig:sio2_ranking}
\end{figure}

Figure~\ref{fig:sio2_ranking} shows that the predicted phase ranking is highly sensitive to dispersion truncation. With direct real-space summation, the ranking fluctuates significantly at standard MLFF cutoffs. Direct summation only converges --- yielding zero pairwise ranking errors --- at an extended interaction cutoff of 54~\AA. Such a large cutoff is computationally prohibitive for on-the-fly evaluation in molecular dynamics.

In contrast, \FastDDD{} recovers the correct polymorph ordering at a modest reciprocal cutoff of $7$~\AA$^{-1}$, at much lower computational cost. This demonstrates that empirical dispersion truncation in MLFFs can introduce qualitative errors in predicted phase stabilities, whereas \FastDDD{} captures the requisite long-range physics efficiently.

\subsection{Water densities}
As a validation in a realistic molecular dynamics setting, we use \FastDDD{} to compute the density of water at 300 K for 1 ns using a water box containing approximately 1500 atoms. Our aim here is to verify that \FastDDD{} reproduces the equilibrium density of a converged D3 treatment, not to assess whether D3 improves agreement with experiment. We compare simulations performed without dispersion corrections and with different D3 interaction cutoff settings to assess their effect on the resulting equilibrium densities. The short-ranged MLFF used in the simulations is a MACE $L = 1$ model trained on a revPBE data set~\cite{o2025towards, advincula2026reactive} without dispersion correction. Details of the model training, simulation setup and convergence studies are provided in Appendix~\ref{sec:app_water_densities}.

Figure~\ref{fig:water_densities} compares the mean densities obtained with different dispersion treatments at 300 K. The model without dispersion correction yields substantially lower densities. In contrast, all dispersion-corrected methods predict significantly higher densities, with values converging to $\sim$ 0.92~g/cm$^3$ with sufficiently large interaction cutoff, consistent with first principle studies~\cite{oneill2026reproducible}.

Among the dispersion-corrected methods, \FastDDD{} reproduces the converged 15~\AA\ D3 reference extremely well. At 300 K, the density differences between \FastDDD{} and the 15~\AA\ cutoff calculations are around 0.2\%, whereas the smaller 6~\AA\ cutoff systematically underestimates the density by approximately 2--3\%. Overall, the close agreement demonstrates that \FastDDD{} accurately reproduces the density obtained from a converged D3 treatment while offering substantially improved computational efficiency.

\begin{figure}[htbp!]
    \centering
    \includegraphics[width=0.85\linewidth]{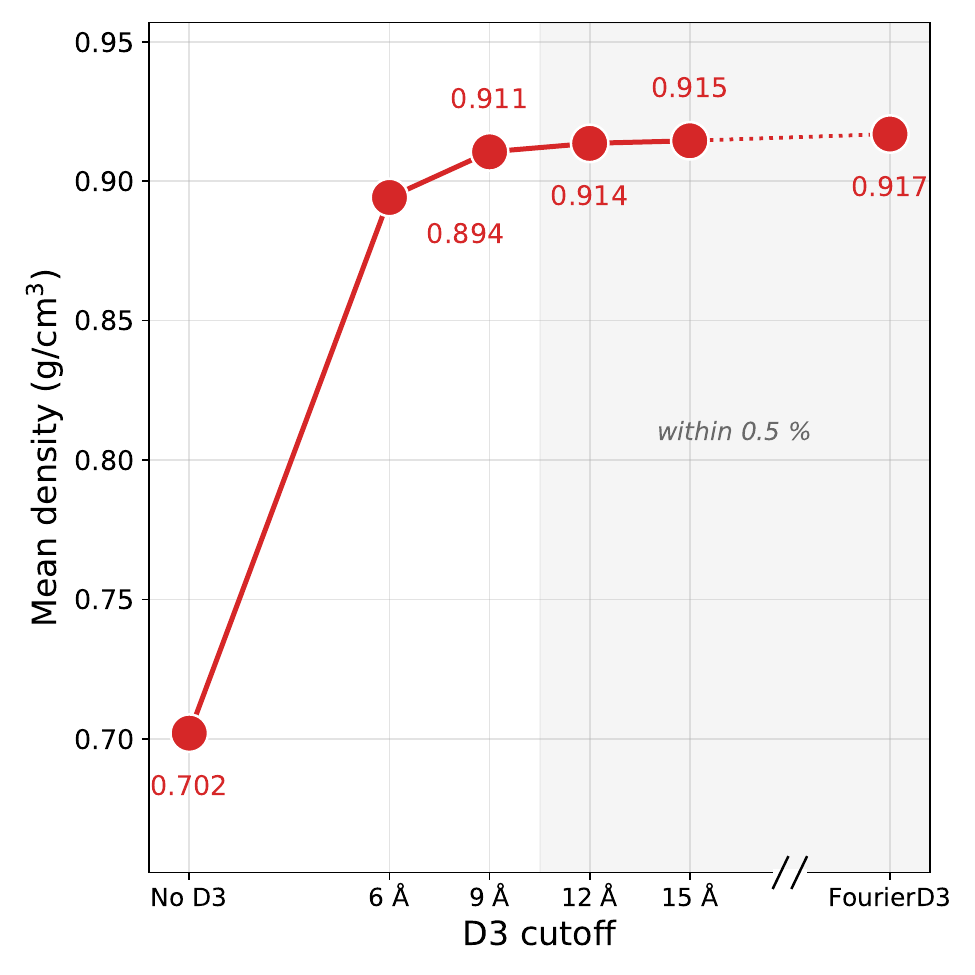}
    \caption{Mean density of liquid water at 300 K for different dispersion treatments. The model without dispersion correction (No D3) predicts substantially lower densities than all dispersion-corrected methods. \FastDDD{} reproduces the converged 15~\AA\ \torchDDD{} reference almost exactly, while a smaller D3 interaction cutoff (6~\AA) underestimates the density.}
    \label{fig:water_densities}
\end{figure}

\subsection{Low-dimensional C structures}
\label{sec:low_d_C_structures}
Having established the accuracy of \FastDDD{} in molecular dynamics simulations through the water density benchmark, we now consider a more controlled test case in which long-range dispersion effects are expected to play a significant role~\cite{chen2013interlayer, qamar2023atomic}.

\begin{figure}[htbp!]
    \centering
    \includegraphics[width=0.95\linewidth]{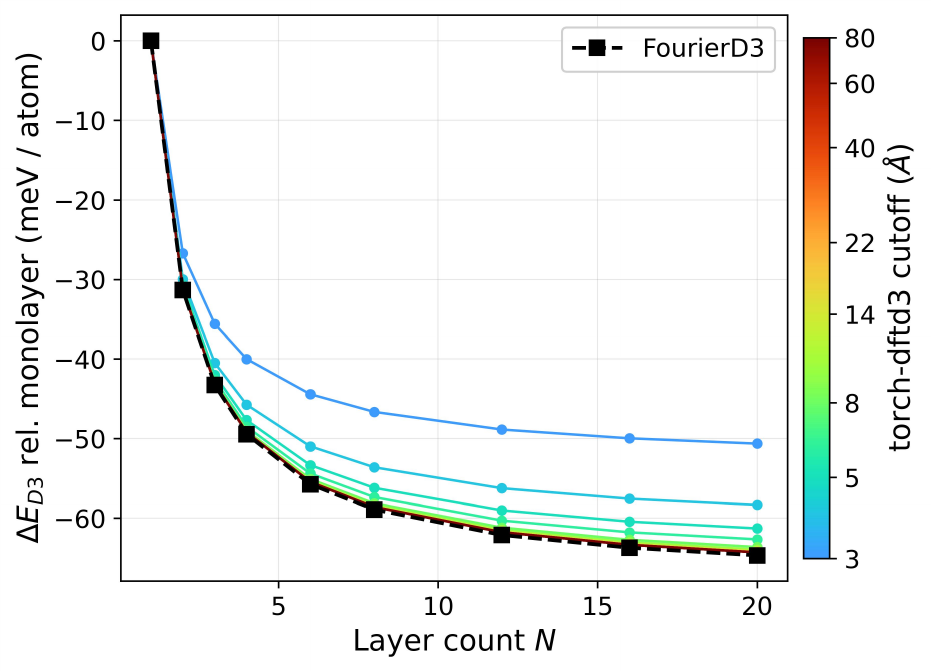}
    \caption{D3 dispersion energy of AA-stacked graphene versus layer number. Coloured curves show \torchDDD{} calculations with real-space cutoffs from $3$ to $80$~\AA, while the black curve denotes the \FastDDD{} Ewald-summed reference ($k_{\rm cut} = 10$~\AA$^{-1}$). Short-range \torchDDD{} systematically underbinds at small cutoffs, whereas larger cutoffs gradually converge toward the \FastDDD{} limit.}
    \label{fig:carbon_layers}
\end{figure}

Figure~\ref{fig:carbon_layers} compares the D3 dispersion energy of AA-stacked $N$-layer graphene computed with \torchDDD{} at several real-space interaction cutoffs against the \FastDDD{} reference. The coloured curves correspond to \torchDDD{} single-point evaluations with cutoffs ranging from $3$ to $80$~\AA, while the black curve denotes the \FastDDD{} result obtained using Ewald summation with a reciprocal-space cutoff of $k_{\rm cut} = 10$~\AA$^{-1}$. System setup details are provided in Appendix~\ref{sec:app_carbon_struct}.

The full-range panel shows that short-range \torchDDD{} systematically underbinds at small cutoffs. The \torchDDD{} curves with interaction cutoffs between $3$ and $10$~\AA\ underestimate the binding energy by roughly $10$~meV/atom relative to the converged result, while a cutoff of $\sim 22$~\AA\ is already visually converged to \FastDDD{} on this scale. The energy difference saturates with increasing layer count $N$, reflecting the finite interaction range imposed by the real-space truncation. Notably, different dispersion correction schemes typically differ by $10$--$20$~meV/atom for layered carbon materials \cite{qamar2023atomic}, indicating that the errors introduced by finite cutoffs can be of a similar magnitude.

This confirms that \FastDDD{} reproduces the long-range, fully converged dispersion contribution at orders-of-magnitude lower cost than a real-space truncation tight enough to reach the same precision.

\section{Summary and Outlook}

We have presented \FastDDD{}, a reformulation of the DFT-D3 dispersion correction that is compatible with reciprocal-space fast summation. The central technical ingredient is a functional tensor decomposition of the environment-dependent $C_6^{\text{ref}}$ reference tensor, which restores the atom-centered separability that particle-mesh methods require. Coupled with a modified coordination number function that admits a well-defined infinite-cutoff limit, this decomposition turns a correction that is typically truncated at short range in MLFF workflows into one that is fully converged --- at a marginal cost.
Our benchmarks on liquid water, liquid benzene, and a seven-element high-entropy alloy show algebraic convergence to a highly-converged \FastDDD-Ewald reference, with a near-constant evaluation time of under 10 ms for systems up to 10,000 atoms. The \ch{SiO2} polymorph analysis demonstrates that this convergence has physical consequences: empirical truncation at standard MLFF cutoffs inverts the predicted ordering of experimentally observed phases, and only the long-range treatment recovers the correct hierarchy.

The mathematical framework is not specific to D3: any pairwise dispersion model whose coefficients depend continuously on atom-centered descriptors — static or dynamic polarizabilities, effective volumes, atomic charges — admits a similar functional tensor decomposition. Extending the same approach to D4~\cite{caldeweyher2017extension, caldeweyher2019generally, caldeweyher2020extension}, Tkatchenko--Scheffler~\cite{tkatchenko2009ts}, and XDM~\cite{becke2007exchangehole} is a natural direction for future work, although each model presents its own structural complications, since each incorporates a different dependence on the atomic environment. For example, DFT-D4 additionally introduces a dependence on charge, while the Tkatchenko--Scheffler, Many-Body Dispersion (MBD), and eXchange--hole Dipole Moment (XDM) models depend on the electron density.

Other fast summation methods~\cite{hardy2009multilevel, hardy2015multilevel, hardy2016multilevel} can be adapted to account for systems with arbitrary boundary conditions. More speculatively, the low-rank decomposition exploited here can be read as a reduced-order representation of a non-local physical quantity in terms of atom-centered features — precisely the kind of representation MLFFs already learn, suggesting opportunities to design MLFF architectures that incorporate non-local physics directly into the learned representation while preserving the scaling advantages of locality.

\section{Statements}

\subsection{Code and Data Availability}
\FastDDD{}{\tt -torch} and the scripts required to reproduce the results are available on GitHub (\url{https://github.com/vicvaleeva/FourierD3}). The accelerated kernel \FastDDD{}{\tt -acc} is available at \url{https://github.com/NVIDIA-BioNeMo/FourierD3}.

\subsection{Acknowledgements}
All computations in this manuscript were performed on an NVIDIA H100 80 GB GPU. We thank the Digital Research Alliance of Canada (\url{alliancecan.ca}) and Dr.\ Jörg Gsponer (Michael Smith Laboratories, University of British Columbia) for providing the computing resources necessary for this work. We thank Dr.\ Christoph Schran and Xavier Advincula for providing the revPBE water dataset.
CO, VV and CH were supported by the Natural Sciences and Engineering Research Council of Canada (NSERC, AWD-017790, AWD-029724) and the Canadian Cancer Society (AWD-031274).

\subsection{Conflicts of Interest}
GC and CO are partners of Symmetric Group LLP that licenses force fields commercially. GC also has equity interest in Angstrom AI.

\subsection{Contributions}
VV and CO conceived the methodology. VV implemented FourierD3{\tt -torch}, the composition-rank scheme, and the SiO{}$_2$ polymorph benchmarks. CH implemented the water and carbon benchmarks. MG, FP and EK developed the accelerated kernel implementation. VV and MG performed the convergence, consistency, and scaling analyses. CO, GC, and EK supervised the work and advised on scientific scope, implementation, and test design. VV, CH, and CO wrote the original draft. All authors reviewed the manuscript and approved the final version.

\section{Appendix}

\subsection{Divergence of the Coordination Number Function}
\label{sec:app_divergence_cn}

\begin{figure}[htbp!]
    \centering
    \includegraphics[width=0.95\linewidth]{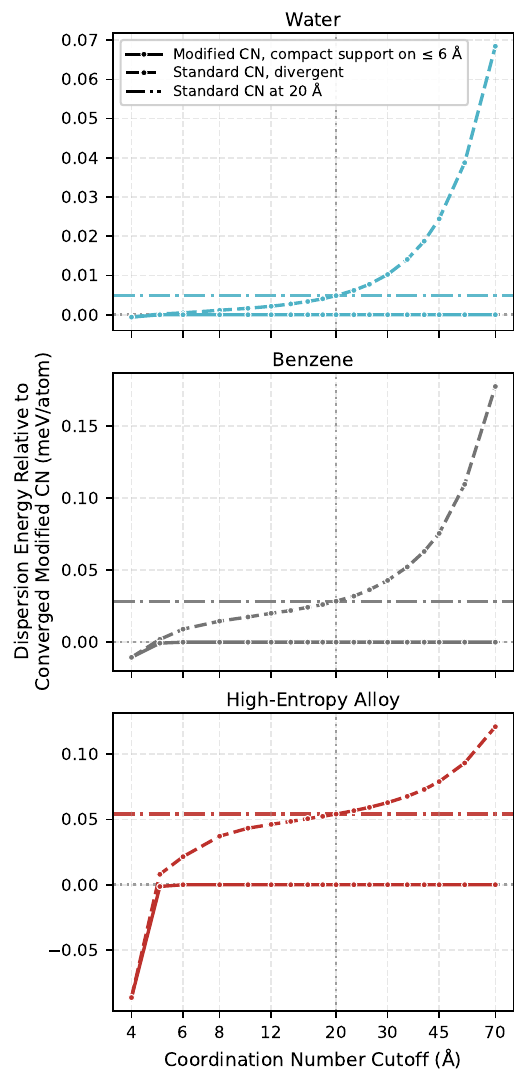}
    \caption{Dispersion energy versus the radius $R_{\rm CN}$ out to which the coordination-number sum is accumulated, relative to the converged value of the modified function of Eq. \ref{eq:D3:smooth_cn_function}. The dispersion sum is evaluated by \FastDDD{} summation, so $R_{\rm CN}$ acts only through the coordination numbers.}
    \label{fig:cn_divergence}
\end{figure}

In the standard DFT-D3 model the pairwise dispersion coefficients depend on the local atomic environment through a continuous coordination number. The counting function used for this purpose does not decay to zero: as $r \to \infty$ every pair contributes $(1+e^{16})^{-1} \approx 1.13 \times 10^{-7}$. The coordination number of an atom in a periodic system is therefore divergent. We stress that this does not render the dispersion energy itself divergent. The reference interpolation of Eq. \ref{eq:coeffs} is normalized, and as $\theta \to \infty$ its weights concentrate on the reference environment with the largest reference coordination number. The interpolated $C^6_{ij}$ therefore remains bounded and approaches a finite value, and, hence, the energy is
also finite at every cutoff. However, this limit is unphysical: the dispersion coefficients of every atom are projected onto a single reference environment irrespective of its actual surroundings, so the environment dependence that motivates the CN construction is lost entirely.

To quantify this behavior we evaluated the dispersion energy as a function of the radius out to which the coordination-number sum is accumulated, for liquid water, liquid benzene and an HEA ($1000$-atom boxes, same setup as in \ref{sec:convergence}). The dispersion lattice sum was evaluated by \FastDDD-Ewald with a $C_6$ decomposition tolerance below the \texttt{float32} numerical floor, so that it is converged independently of $R_{\rm CN}$. Figure~\ref{fig:cn_divergence} reports the resulting energies relative to the converged value of the modified CN function (\ref{eq:D3:smooth_cn_function}). With the standard function the energy difference keeps rising: extending the cutoff from the conventional $20$~\AA\ to $45$~\AA\ shifts it by $0.020$, $0.047$ and $0.025$~meV/atom for water, benzene and the HEA respectively, and it is still changing monotonically at $R_{\rm CN} = 70$~\AA. With the modified function the energy is unchanged at every radius beyond its $6$~\AA\ support. The eventual limit predicted above is not visible in this window, since it is only achieved at several hundred Angstrom. We note that the energy difference between the reference CN function with the $20$~\AA\ cutoff and our modified CN function (Eq. \ref{eq:D3:smooth_cn_function}) is similar to the difference between the reference CN function truncated at $20$~\AA\ and $\sim 50-70$~\AA.

\subsection{Decomposition of the Damping Function}
\label{sec:decomp_damp}
Both classes of damping functions can be written in the form 
\[
  f^{\rm damp}(r, Z_i, Z_j) = \frac{r^\alpha}{r^\alpha + (a_1 R_{ij} + a_2)^\alpha},
\]
where $\alpha$ is a positive integer. First, we choose a normalization length $\bar{R}$, a representative length for $R_{ij}$, and non-dimensionalize via $s = r/\bar{R}$, $S_{ij} = R_{ij} / \bar{R}$, $b_1 = a_1$, $b_2 = a_2 / \bar{R}$, to obtain
\[
    f^{\rm damp}(r, Z_i, Z_j) = 
    \frac{s^\alpha}{s^\alpha + (b_1 S_{ij} + b_2)^\alpha}.
\]
Next, we substitute $t = 1/(1+s)$, i.e.\ $s = 1/t - 1$, to rewrite the damping function as
\begin{equation*}
\begin{aligned}
    f^{\rm damp}(r, Z_i, Z_j) &= g^{\rm damp}(t, S_{ij}) \\
    &= \frac{(1-t)^\alpha}{(1-t)^\alpha + t^\alpha (b_1 S_{ij} + b_2)^\alpha}.
\end{aligned}
\end{equation*}
These transformations make $g^{\rm damp}$ a smooth function on the compact domain $(t, S_{ij}) \in [0, 1] \times [S_0, S_1]$ and therefore guarantee a rapidly converging Chebyshev expansion, which can be readily transformed into
\[
    f^{\rm damp}(r, Z_i, Z_j) = g^{\rm damp}(t, S_{ij})
 \approx \sum_{\ell = 0}^{\ell_{\rm max}} S_{ij}^\ell g_\ell(t).
\]
Since $S_{ij}$ is separable in $Q_{Z_i} Q_{Z_j}$, this provides the desired decomposition. The precise rank $\ell_{\rm max}$ depends on the specific damping function and the desired accuracy. It is clear, however, that the number of terms depends only on the parameters $a_1, a_2, \alpha$ and the range of $S_{ij} = R_{ij} / \bar{R}$, not on the number of chemical elements. As noted above, this decomposition introduces a non-trivial Fourier transform of the potential, which precludes a fast practical implementation of the method.

\subsection{\ch{SiO2} Structures List}
\label{sec:sio2_data}
A list of IDs of silicon dioxide polymorph structures from the Materials Project~\cite{mp1, mp2}:
{\tt mp-1199998, mp-1200292, mp-1203655, mp-17279, mp-18280, mp-546794, mp-554267, mp-554755, mp-555521, mp-556654, mp-557894, mp-558115, mp-558351, mp-560155, mp-600080, mp-615993, mp-644923, mp-669426, mp-680204, mp-6947, mp-733790, mp-10948, mp-1188220, mp-1195265, mp-1205213, mp-16964, mp-32895, mp-553945, mp-555891, mp-556218, mp-556454, mp-556469, mp-558025, mp-560809, mp-560826, mp-561090, mp-561181, mp-561351, mp-7648, mp-8352, mp-9258, mp-972808, mp-1071820, mp-15078, mp-555235, mp-556591, mp-557881, mp-557933, mp-558326, mp-559872, mp-560336, mp-560920, mp-560941, mp-560973, mp-560998, mp-600037, mp-600054, mp-640556, mp-653763, mp-667368, mp-667448, mp-6945, mp-7905, mp-10851, mp-11684, mp-1201887, mp-1204057, mp-12787, mp-17909, mp-542814, mp-556262, mp-556961, mp-559091, mp-559550, mp-560064, mp-640917, mp-646895, mp-651707, mp-6922, mp-6930, mp-7000, mp-7087, mp-8059, mp-8602}

\subsection{Water Density Simulations}
\label{sec:app_water_densities}
A new MACE model was trained on a water dataset computed at the revPBE level of theory. No D3 dispersion correction was included in the dataset, but dispersion was added at inference time using the D3 implementations evaluated in this benchmark. The model used hidden irreducible representations \texttt{128x0e+128x1o}, a correlation order of $3$, a maximum message angular momentum of $l_{\max}=3$, and a radial cutoff of $6 $ \AA{} with two layers of message passing. Training was performed using the universal loss and the Adam  optimiser with AMSGrad, with isolated-atom reference energies supplied. On the validation set, the resulting model achieved energy and force RMSEs of $2.2$ meV/atom and $34.5$ meV/\AA, respectively.

Water-density NPT simulations were performed using the Atomic Simulation Environment (ASE) with periodic boundary conditions. A cubic box containing 512 water molecules (1536 atoms) with an initial side length of $24.84$~\AA\ was used. Simulations were carried out at 300~K using a Berendsen NPT scheme, with a temperature-coupling time of $0.1$~ps and a pressure-coupling time of $1$~ps at $1.01325$~bar.

The system was simulated for $2 \times 10^6$ steps with a timestep of 0.5~fs, corresponding to a total simulation time of 1~ns. The first 200~ps were discarded as equilibration, and the remaining trajectory was used to compute the average density.

Each density trajectory was sampled every 0.1 ps. The plotted band shows the autocorrelation-corrected standard error of the mean, estimated from an FFT-based autocorrelation function using blocks of 10 samples.

\subsection{Low-dimensional C structures}
\label{sec:app_carbon_struct}
AA-stacked graphene structures were constructed by stacking identical graphene layers along the $z$-axis. Each layer contains two carbon atoms arranged on a hexagonal lattice with in-plane lattice parameter $a = 2.46$~\AA. The interlayer spacing was fixed at $3.35$~\AA. Structures containing up to 20 layers were considered. The benchmark focuses on the size dependence of the dispersion energy for a fixed reference geometry. Structural optimization was not pursued further due convergence challenges associated with large dispersion cutoffs.

For the \FastDDD{} evaluation, a vacuum region of 100~\AA\ was added along the stacking direction to avoid interactions between periodic images, while retaining the three-dimensional periodic boundary conditions required by the \FastDDD{} Ewald summation implementation.

\subsection{NPT Benchmarks}
\label{sec:bench}

To assess the cost of the dispersion correction in production molecular dynamics, we ran Berendsen NPT simulations (ASE; $T = 300$\,K, $p = 1.01325$\,bar,$\tau_T = 0.1$\,ps, $\tau_p = 1$\,ps, $\Delta t = 0.5$\,fs) with MACE (cuEquivariance, float32) plus D3(BJ), evaluated either with \FastDDD{} or with the highly-optimized GPU real-space code LASP-D3\cite{Luo2026-ni}. We considered two systems: liquid water (4913 molecules, 14{,}739 atoms; revPBE MACE model from \ref{sec:app_water_densities}) and an Al--Cu--Ag--Au--Ni--Pd--Pt high-entropy alloy (16{,}384 atoms; \texttt{mace-omat-medium}). \FastDDD{} used a $64^3$ mesh with order-6 B-splines in float32, and reused MACE's 6\,\AA{} neighbour list for the coordination numbers. LASP-D3 used dispersion/coordination-number cutoffs of 12/12\,\AA{} for water and 40/20\,\AA{} for the alloy. The short cutoffs for water allow LASP-D3 to use an efficient cell list implementation, but large cutoffs required for the HEA would require immensely large simulation cells, hence, LASP-D3 falls back to the naive neighbour list implementation.
Throughput was measured over 5000 steps after a 2000-step warm-up on a single NVIDIA H100 (80\,GB). For water, the MACE+\FastDDD{} approach performs only slightly faster (51.6 vs.\ 50.8\,ps/day; 836 vs.\ 851\,ms/step), since the MACE evaluation dominates the cost. For the alloy, the long cutoffs required by LASP-D3 significantly decrease the combined performance: MACE+\FastDDD{} reaches 56.5\,ps/day against 34.2\,ps/day for MACE+LASP-D3 (765 vs.\ 1264\,ms/step).

\subsection{Low-rank Decomposition: Energies}
\label{sec:c6tol}

The rank $R$ is selected in Sec.~\ref{sec:results:decomp} by requiring the maximum relative error $\varepsilon_{\max}(R)$ of Eq. \ref{eq:decomp} to fall below a fixed tolerance. Here we numerically confirm that this criterion controls the energy. For each reference system, same as in \ref{sec:convergence}, we evaluate the dispersion energy at every rank with all remaining settings held fixed, in \texttt{float64}.

\begin{figure}[htbp!]
    \centering
    \includegraphics[width=0.95\linewidth]{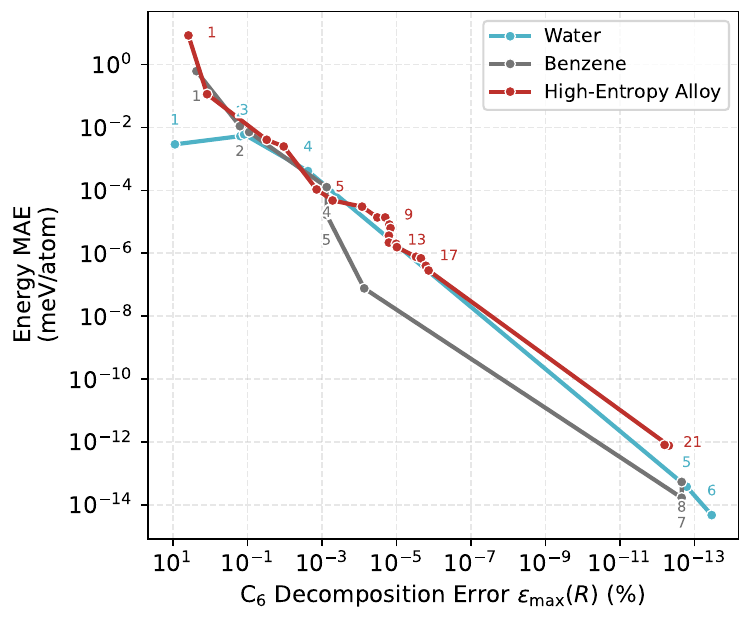}
    \caption{Energy error against the accuracy of the low-rank decomposition, marker denotes the rank $R$. The x-axis is $\varepsilon_{\max}(R)$, decreasing to the right; the y-axis is the mean absolute energy error against a decomposition exact to machine precision. The error reaches the float64 round-off of the Ewald summation once the
    decomposition is exact.}
    \label{fig:c6tol}
\end{figure}

\bibliography{references}

\end{document}